\documentstyle[preprint,aps,epsfig]{revtex}

\begin{document}
\newcommand{\lsim}{\mathrel{\mathop{\kern 0pt \rlap
  {\raise.2ex\hbox{$<$}}}
  \lower.9ex\hbox{\kern-.190em $\sim$}}}
\newcommand{\gsim}{\mathrel{\mathop{\kern 0pt \rlap
  {\raise.2ex\hbox{$>$}}}
  \lower.9ex\hbox{\kern-.190em $\sim$}}}
\newcommand{\es}     {\epsilon}
\newcommand{\Dt}     {\Delta}
  \newcommand{\no}     {\nonumber}
  \newcommand{\gm}     {\gamma}
  \newcommand{\Gm}     {\Gamma}
\newcommand{\lm}     {\lambda}
\newcommand{\sg}     {\sigma}
\newcommand{\Lm}     {\Lambda}
\newcommand{\beq}     {\begin{equation}}
\newcommand{\eeq}     {\end{equation}}
\newcommand{\bea}     {\begin{eqnarray}}
\newcommand{\eea}     {\end{eqnarray}}

\draft

\preprint{
\vbox{
 \hbox{ {\bf hep-ph/0103127},~~YUMS-01-01,~~KIAS P01014}
}}

\title{Muon Anomalous Magnetic Moment $(g-2)_\mu$ \\
and the Randall-Sundrum Model}

\author{
C.~S. Kim$^{1,}$\footnote{kim@kimcs.yonsei.ac.kr},~~
J.~D. Kim$^{1,}$\footnote{jdkim@theory.yonsei.ac.kr}~~ and~~
Jeonghyeon Song$^{2,}$\footnote{jhsong@kias.re.kr}
}

\address{$^1$Department of Physics and IPAP,
 Yonsei University, Seoul 120-749, Korea\\
 $^2$School of Physics,
Korea Institute for Advanced Study, Seoul 130-012, Korea}

\vskip -0.75cm

\maketitle
\begin{abstract}
\vskip-5ex 

\noindent
We study the effects of the Kaluza-Klein gravitons
in the Randall-Sundrum model
on the recent BNL measurements of the
muon $(g-2)$ deviation from the standard model prediction.
By examining the $J$-partial wave unitarity bounds
of the elastic process $\gm\gm\to\gm\gm$,
the cut-off on the number of massive KK gravitons, $n_c$,
has been introduced.
We found that the recently measured $\Dt a_\mu$
can be accommodated in the RS model,
within the natural parameter space
allowed by the perturbative unitarity.
For example, dozens (hundreds) of the $n_c$
for $\Lm_\pi=1\sim2$ TeV (3 TeV)
can explain the reported $\Dt a_\mu$.
\end{abstract}

\newpage


\noindent
{\bf 1. Introduction.}~~~
As one of the most accurately measured quantities
in particle physics,
the anomalous magnetic moment of the muon,
$(g-2)_\mu$,
has played an important role in providing
constraints on various models for physics
beyond the standard model (SM).
The recent Brookhaven E821 results for the $a_\mu \equiv (g-2)_\mu/2$
by a 2.6 standard-deviation
from its SM prediction\cite{Brown:2001mg},
\beq
\label{Dta}
\Dt a_\mu = a_\mu ({\rm exp})-a_\mu ({\rm SM}) =
(43 \pm 16) \times 10^{-10},
\eeq
therefore,
is very likely to indicate that
the SM should be extended just above the electroweak scale.
A conservative view still exists such that
the uncertainties in the SM calculations
of the hadronic vacuum polarization
may explain the deviation\cite{Yndurain:2001qw}.
The implications of the $\Dt a_\mu$
on new physics have been extensively studied.
Supersymmetric models are shown
to accommodate  this $\Dt a_\mu$,
as being consistent
with all other relevant constraints\cite{susy}.
Interesting results are that the reported $\Dt a_\mu$ value
favors large $\tan\beta$ case
and imposes observable upper bounds on the masses
of supersymmetric particles.
The effects from other new physics such as
technicolor models\cite{Xiong:2001rt},
lepton flavor violation\cite{Ma:2001mr},
composite models\cite{Das:2001it},
leptoquark models\cite{Cheung:2001ip},
and others\cite{Choudhury:2001ad}
have been also studied.

Based on a theoretical motivation
to explain the gauge hierarchy problem,
a novel approach has been proposed by introducing extra dimensions:
Arkani-Hamed, Dimopoulos and
Dvali (ADD) proposed that
the large volume of the extra dimensions
with factorizable geometry can explain the
observed largeness of Planck scale $M_{\rm Pl}$,
whereas the fundamental string scale
$M_S$ being maintained around TeV scale{\cite{ADD}};
based on two branes and a single extra dimension
with non-factorizable geometry,
Randall and Sundrum (RS) have proposed another
high dimensional scenario, where
the gauge hierarchy is attributed to a geometrical
exponential factor{\cite{Randall:1999ee}}.
The phenomenologies of both extra-dimensional models
have been extensively studied\cite{phenom}.
In this paper, we examine if the RS model can explain the
observed $\Dt a_\mu$ deviation.

In the four-dimensional effective theory,
both extra dimensional
scenarios yield the Kaluza-Klein (KK)
states of massive gravitons.
The KK gravitons in the ADD case have
almost continuous spectrum
up to a cut-off scale $M_S$
with the Planck suppressed interaction;
those in the RS case
show discrete mass spectrum about TeV scale
with the electroweak scale $(\Lm_\pi)$ interactions.
In the ADD case,
the calculation of the KK graviton contribution to the muon
$(g-2)$ has shown some interesting results such
that a given KK graviton generates finite contribution to the $\Delta
a_\mu$\cite{Graesser:2000yg};
the universal coupling of each KK
graviton leads to cancellation
among logarithmic divergences of each
Feynman diagram;
the contribution of the KK graviton much heavier
than the muon shows non-decoupling.
The finiteness of the
contributions of a graviton has been well known in
supergravity models\cite{delAguila:1984fv}.
In the ADD model, the
summation of all the KK graviton contributions
with the cut-off $M_S$
is compensated by the Planck scale suppression,
which leaves finite contribution
to the $\Dt a_\mu$ proportional to $(m_\mu/M_S)^2$.

In the RS model, there exist an infinitely
large number of KK graviton
states in the view point of our four-dimensional world,
whereas the interaction is suppressed by only the
electroweak scale.
A naive summation of the finite contributions
shall generate infinite contribution to the $\Dt a_\mu$.
A cut-off, corresponding to the $M_S$ in the ADD case,
is needed.
We introduce $n_c$,
the unitarity cut-off on the number of the KK graviton states.
Then, we express the $\Dt a_\mu$
in terms of $n_c$ and $\Lm_\pi$.
In fact,  the necessity of introducing a cut-off
in the RS model has been already known
in its phenomenological study:
Unitarity violation happens
at high energies\cite{Davoudiasl:2000jd}.
By examining the $J$-partial wave amplitudes of the
elastic process $\gm\gm\to\gm\gm$,
we shall derive the perturbative unitarity constraints
on $n_c$ and $\sqrt{s}/\Lm_\pi$.
It is to be shown that
the recently measured $\Dt a_\mu$
can be accommodated in the RS model,
with the natural values of $n_c$ and $\Lm_\pi$
allowed by the unitarity constraint.
\\

\noindent
{\bf 2. Randall-Sundrum Model and $\Dt a_\mu$.}~~~
For the hierarchy
problem, Randall and Sundrum have proposed a five-dimensional
non-factorizable geometry with the extra dimension
compactified on a $S_1/Z_2$ orbifold
of radius $r_c$\cite{Randall:1999ee}.
Reconciled with four-dimensional
Poincar${\rm \acute{e}}$ invariance,
the RS configuration has the following solution to
Einstein's equations:
\begin{equation}
\label{metric}
d s^2=
e^{-2 k r_c |\varphi| }
\eta_{\mu\nu} d x^\mu d x^\nu
-
r_c^2 d \varphi^2,
\end{equation}
where $0 \leq |\varphi| \leq \pi$,
and $k$ is $ AdS_5$
curvature.
Two orbifold fixed points accommodate two three-branes,
the hidden brane at $\varphi=0$ and
our visible brane at $|\varphi|=\pi$.
The arrangement of our brane at $|\varphi|=\pi$
renders a fundamental scale $m_0$ to appear as
the four-dimensional physical mass $m=e^{-k r_c \pi} m_0$.
The hierarchy problem can be answered
if $k r_c \approx 12$.
From the four-dimensional effective action,
the relation between the four-dimensional
Planck scale $M_{\rm Pl}$
and the fundamental string scale $M_5$
is obtained by
$M_{\rm Pl}^2={M_5^3}(1-e^{-2kr_c \pi})/k$.

The compactification of the fifth dimension
leads to the following four-dimensional
{\it effective} Lagrangian
\cite{Davoudiasl:2000jd},
\begin{equation}
\label{Lgraviton}
{\mathcal L} = -\frac{1}{M_{\rm Pl}}
T^{\mu\nu} h^{(0)}_{\mu\nu}
-\frac{1}{\Lambda_\pi}
T^{\mu\nu} \sum_{n=1}^\infty
h^{(n)}_{\mu\nu}
\,,
\end{equation}
where $ \Lambda_\pi \equiv  e^{-k r_c \pi} M_{\rm Pl}$.
The coupling of the zero mode KK graviton
is suppressed by usual Planck scale $M_{\rm Pl}$,
while those of the massive KK gravitons
by the electroweak scale
$\Lambda_\pi$.
The masses of the KK gravitons are also at electroweak scale,
$$
m_n= x_n ({k}/{M_{\rm Pl}} ){\Lambda_\pi},
$$
where $x_n$ is  the  $n$-th root
of the first order Bessel function.
And the total decay width of the $n$-th KK graviton
is
$\Gamma_n = \rho \,m_n x_n^2 (k/M_{\rm Pl})^2 $,
where the $\rho$, fixed to be one,
is a model-dependent parameter
which is different according to the decay
pattern of the KK graviton.
The condition $k<M_{\rm Pl}$ is to be imposed
to maintain the reliability of the RS solution
in Eq.~(\ref{metric})\cite{Davoudiasl:2000tf}.

The Lagrangian in Eq.~(\ref{Lgraviton})
then makes new contributions
to the $\Dt a_\mu$.
The relevant Feynman diagrams mediated by the KK gravitons
are presented in Fig.~1.
We refer readers to Ref. \cite{feynman}
for the corresponding
Feynman rules of the graviton, photon and
fermion.
Note that the gravitational Ward identity
renders almost negligible, to leading order,
the contributions of the terms containing
two or more $k_\alpha$'s
in the massive graviton propagator.
Here the $k_\alpha$ denotes the momentum of the graviton.
A given $n$-th KK graviton state generates the $\Dt a_\mu$,
which is parameterized by \cite{Graesser:2000yg}
\beq
\Dt a_\mu^{(n)}
=\frac{1}{16 \pi^2}
\left(
\frac{m_\mu}{\Lm_\pi}
\right)^2
\sum_f a_f^G
= \frac{5}{16 \pi^2}
\left(
\frac{m_\mu}{\Lm_\pi}
\right)^2
\,,
\eeq
where the $f$ runs over five Feynman diagrams
and the second equality holds for the case of $m_n \gg m_\mu$.
As discussed in the previous section,
naive summation over all the KK states
would yield infinite contribution, i.e.,
$\Dt a_\mu \equiv \sum_n^\infty \Dt a_\mu^{(n)} \to \infty$.
A cut-off is needed.

This requirement of a cut-off is expected since
the RS model described by Eq.~(\ref{Lgraviton}) is
an effective theory.
According to the analysis of the process
$e^+ e^- \to \mu^+ \mu^-$ with the KK graviton effects,
in the large $k/M_{\rm Pl}$ case
the partial wave unitarity seems to be violated
even at TeV scale as
the resonant peaks of the KK gravitons
become very wide and the cross section
is greatly enhanced \cite{Davoudiasl:2000jd}.
Since in the small $k/M_{\rm Pl}\lsim 0.3$ case
the unitarity appears preserved at several TeV scales,
possible unitarity violation of the RS model
at high energy colliders
has not been considered seriously yet.
However,
when we consider the loop effects by using an effective
theory (the RS model) with an infinitely large number of states,   we should
determine the cut-off scale below which the perturbative calculations are
reliable, and  add only the contributions
up to the cut-off scale.
We introduce the cut-off on the effective number
of KK graviton states,
$n_c$.
Then, the total RS contribution to the $\Dt a_\mu$
becomes
\beq
\Dt a_\mu ^{RS}
=\frac{5}{16 \pi^2}
\left(
\frac{m_\mu}{\Lm_\pi}
\right)^2\, n_c.
\eeq
The scale of $n_c$ will be estimated
by examining the partial wave unitary bound
in the process of $\gm\gm\to\gm\gm$.
Note that the parameters $(\Lm_\pi, k/M_{\rm Pl})$
are constrained from the phenomenological studies of
virtual exchange of the KK gravitons
at LEP II and Tevatron Run I data;
$\Lm_\pi \gsim 1.1 ~(0.65)$ TeV for the
$k/M_{\rm Pl} =0.1 ~(0.3)$.

In Fig.~2, we plot the $\Dt a_\mu^{RS}$ as a function of $n_c$
for  $\Lm_\pi=$1, 2, 3 TeV.
It can be seen that  the RS model
can explain the recently reported $\Dt a_\mu$
from the SM prediction with $n_c\simeq  10$, 50, and 150,
for $\Lm_\pi=1 $, 2, and 3 TeV, respectively.
As shall be shown in the next section,
the scale of $n_c \simeq 10-100$
is natural in the sense that
the partial wave unitarity is well preserved
in this range.
The scale of the heaviest KK graviton mass,
$m_{n_c}$,
then becomes of order 10 TeV,
which
can be regarded as a naive cut-off scale of the
four-dimensional effective Lagrangian
in the RS scenario.
In the next section, we will show that
this is consistent
with the cut-off scale for the perturbative unitarity.

Some discussions  on the other RS effects
rather than the KK gravitons are in order here.
First in the RS scenario where
the modulus field
(called the radion)
is inherent,
its stabilization is needed for the consistency
in a cosmological context \cite{Giudice:2000av}.
According to the Goldberger and Wise stabilization mechanism,
this radion can be an order of magnitude lighter than the
$\Lm_\pi$ \cite{Giudice:2000av,Ko,Goldberger:1999uk}.
The radion contributions to the $\Dt a_\mu$,
however, have been too small to explain alone the
$\Dt a_\mu$ of $\sim 4 \times 10^{-9}$\cite{radion-g}.
The effects of an extended RS model,
where the SM gauge and fermion fields,
are both in the bulk have been also
examined\cite{Davoudiasl:2000my}.
The dominant contributions from bulk gauge bosons,
bulk fermions,
and Higgs bosons, however, generate
$negative$ contribution to the $\Dt a_\mu$,
which is difficult to be compatible
with the recently reported positive deviation in $a_\mu$.
Moreover, the graviton loop contributions
are not well-defined
since the cancellation among logarithmic divergences
does not occur due to non-universal couplings
between the KK gravitons and the bulk SM fields.
\\

\noindent
{\bf 3. Unitary Bound from $\gm\gm\to\gm\gm$
Scattering.}~~~
In order to see whether the $n_c\simeq 10-100$
can naturally explain the $\Dt a_\mu$
in the effective theory interpretation,
we consider another process
to estimate the scale of $n_c$.
The elastic process
$\gamma \gamma \to \gamma\gamma$ is to be examined,
focused on the unitarity bounds
of the RS model:
It has some merits such that
the RS effects mediated by KK gravitons are dominant
due to the absence of the SM contributions
at tree level
and thus the unitarity bounds of the RS model can be more
sensitively probed
compared to other processes.
The requirement of partial wave unitarity
shall constrain $\sqrt{s}/\Lm_\pi$
and the number of the KK graviton states, $n_c$.
The $J$-partial wave amplitude is defined \cite{Eboli:2000aq} by
\begin{equation}
a^J_{\mu\mu^\prime} = \frac{1}{64 \pi}
\int^1_{-1} d\cos\theta\; d^J_{\mu\mu^\prime}(\cos\theta)
\,\left[
-i{\cal M}_{\lambda_1\lambda_2\lambda_3\lambda_4}\right] \; ,
\label{aj}
\end{equation}
where
the ${\cal M}_{\lm_1\lm_2\lm_3\lm_4}$
is the helicity amplitudes,
$\mu =\lambda_1-\lambda_2$,
$\mu^\prime = \lambda_3-\lambda_4$,
and the $d^J_{\mu\mu^\prime}$ is the Wigner functions\cite{pdg}.
Unitarity implies that
the largest
eigenvalue ($\chi$)
of $a^J_{\mu\mu^\prime}$
is to be $|\chi| \le 1$.
The reliability of perturbative calculations
is approximately guaranteed
by the conditions
$|\chi|= 1$ and $|{\rm Re}(\chi )| = 1/2$.
The helicity amplitudes,
of which the dominant contribution at high energies
comes from the KK gravitons,
are
\begin{eqnarray}
{\cal M}_{++++} &=& {\cal M}_{----}
= -i~\frac{s^2}{\Lm_\pi^2}~
\sum_{n=1}^{n_c}\left[
D_n(t)
+
D_n(u)
\right]\; , \\ \no
{\cal M}_{+-+-} &=& {\cal M}_{-+-+}
= -i~\frac{u^2}{\Lm_\pi^2}~
\sum_{n=1}^{n_c}\left[
D_n(s)
+
D_n(t)
\right]\;, \\ \no
{\cal M}_{+--+} &=& {\cal M}_{-++-}
= -i~\frac{t^2}{\Lm_\pi^2}~
\sum_{n=1}^{n_c}\left[
D_n(s)
+
D_n(u)
\right]\;,
\end{eqnarray}
where $D_n(s)=1/(s-m_n^2+i m_n \Gm_n)$
with $m_n$ being the mass of the $n$-th KK graviton
state.
Note that two parameters $(\sqrt{s}/\Lm_\pi, n_c)$
with a given $k/M_{\rm Pl}$ determine
the helicity amplitudes,
and thus $\chi$.
The odd $J$-partial wave amplitudes vanish
due to Bose-Einstein statistics
in the elastic $\gamma \gamma$ scattering.
And we have $a^2_{22} = a^2_{-2-2}$ and
$a^2_{-22} = a^2_{2-2}$ from the parity
arguments.
The non-vanishing  eigenvalues $\chi_i$ are
$a^2_{00}$ and $2 \, a^2_{22}$.

In two limiting cases where $\sqrt{s}/\Lambda_{\pi} \ll
(k/M_{\rm Pl})x_n$ and $\sqrt{s}/\Lambda_{\pi} \gg
(k/M_{\rm Pl})x_n$,
the two non-vanishing eigenvalues show converging behaviors.
When  $a^{J(n)}_{\mu\mu'}$ denotes the contribution
of the $n$-th KK state to $a^J_{\mu\mu'}$,
the $a^{2(n)}_{00}$ is
approximated as
\begin{eqnarray}
a^{2(n)}_{00}  &\longrightarrow  &
\frac{1}{384\pi}\frac{s^4}{\Lambda^8_{\pi}}
\left(\frac{k}{M_{\rm Pl}} \right)^{-6}
\frac{1}{x_n^6} , \quad \quad\quad\qquad\qquad ~ ~
{\mbox{ for }}  \sqrt{s}/\Lambda_{\pi} \ll
(k/M_{\rm Pl})x_n,
\\ \nonumber
a^{2(n)}_{00}  &\longrightarrow&\frac{1}{16\pi}\frac{s}{\Lambda^2_{\pi}}
\left[\log\left\{2 \frac{k^2}{M_{\rm Pl}^2}
\frac{\Lambda^2_{\pi}}{s}x^2_n\right\}+3\right],\quad\qquad
{\mbox{ for }}\sqrt{s}/\Lambda_{\pi} \gg
(k/M_{\rm Pl})x_n.
\end{eqnarray}
There exist finite unitarity bounds $n_c \lsim 1000$ at small
$\sqrt{s}/\Lambda_{\pi}$, depending on the value of the parameter
$k/M_{\rm Pl}$.  Similarly, we calculate the
approximate formula for $a^{2(n)}_{22}$ in two limiting cases:
\begin{eqnarray}
a^{2(n)}_{22}  &\longrightarrow&
-\frac{1}{160\pi}\frac{s^2}{\Lambda^4_{\pi}}\frac{M^2_{Pl}}{k^2}
\frac{1}{x_n^2}, \qquad\qquad\qquad\qquad\qquad~ {\mbox{ for }}
\sqrt{s}/\Lambda_{\pi}
\ll (k/M_{\rm Pl})x_n
,  \\   \nonumber
a^{2(n)}_{22}  &\longrightarrow & -\frac{1}{64\pi}\frac{s}{\Lambda^2_{\pi}}
\left[\frac{1}{16}\log\left\{2\frac{k^2}{M_{\rm Pl}^2}
\frac{\Lambda^2_{\pi}}{s} x^2_n\right\}+\frac{2}{5}\right], ~~~ \;\;\; {\mbox{
for }}
\sqrt{s}/\Lambda_{\pi}
\gg
(k/M_{\rm Pl})x_n
.
\end{eqnarray}
The unitary bounds $n_c$ are
around $n_c\simeq 10\sim 100$.
The eigenvalue $2 a^2_{22}$ generates
stronger constraints compared to that of $a^2_{00}$.

Figures 3 and 4 exhibit the exact numerical calculations
of the unitarity bounds on
$(\sqrt{s}/\Lm_\pi, n_c)$ plane from the $\chi=a^2_{00}$  and
$\chi=2\,a^2_{22}$ constraints, respectively.
We consider three values for $k/M_{\rm Pl}(=$
0.1, 0.3, 0.7).
Note that the upper bounds on
$\sqrt{s}/\Lm_\pi$
constrained from the perturbative unitarity
are almost the same
in the range of $n_c \simeq 10 - 100$
(such that $\sqrt{s}\lsim 4\,\Lm_\pi$, $6.5\, \Lm_\pi$,
and $11\,\Lm_\pi$
for $k/M_{\rm Pl}=0.1$, 0.3, and 0.7 cases,
respectively).
When we regard this upper bounds on $\sqrt{s}$
as the cut-off scale of the
effective RS model,
we can see that
this cut-off scale
is compatible with the scale of $m_{n_c}$.
Thus,
the cut-off on the number of KK graviton states
$( n_c)$
which explains the reported $\Dt a_\mu$
is within the allowed region
by the perturbative unitarity bounds.
\\

\noindent
{\bf 4. Summary.}~~~
To summarize, we have studied the effects of the
Kaluza-Klein gravitons
in the Randall-Sundrum model
on the recent BNL measurements of the
muon $(g-2)$ deviation
from the SM prediction.
It is known that the individual contribution of a heavy
KK graviton to $\Dt a_\mu$ is finite.
Since the four-dimensional effective theory in the RS
model contains an infinitely large number of
massive KK gravitons with the TeV suppressed couplings,
a naive summation would yield infinite $\Dt a_\mu$.
A cut-off on the number of the KK gravitons
has been introduced.
Then the reported $\Dt a_\mu$ can be attributed to the RS effects
with dozens (hundreds) of the $n_c$
for $\Lm_\pi=1\sim2$ TeV (3 TeV).
By examining
the $J$-partial wave unitarity bounds
in the elastic process $\gm\gm\to\gm\gm$,
we have shown that
this range of $n_c$ is natural
in the effective theory interpretation
and the scale of the heaviest graviton mass
$(m_{n_c})$ is compatible with the cut-off scale
for
the perturbative unitarity.
Therefore, we conclude that the recently measured $\Dt a_\mu$
can be accommodated in the RS model.
\\

\noindent
{\bf Acknowledgments.}~~~
The work of C.S.K. was supported
in part by  BK21 Program, SRC Program
and Grant No. 2000-1-11100-003-1
of the KOSEF, and in part by the KRF Grants,
Project No. 2000-015-DP0077.
The work of J.D.K was supported
by the Korea Research Foundation Grants
(KRF-2000-DA0013 and 2000-015-DP0077).



\begin{center}
\begin{figure}[htb]
\hbox to\textwidth{\hss\epsfig{file=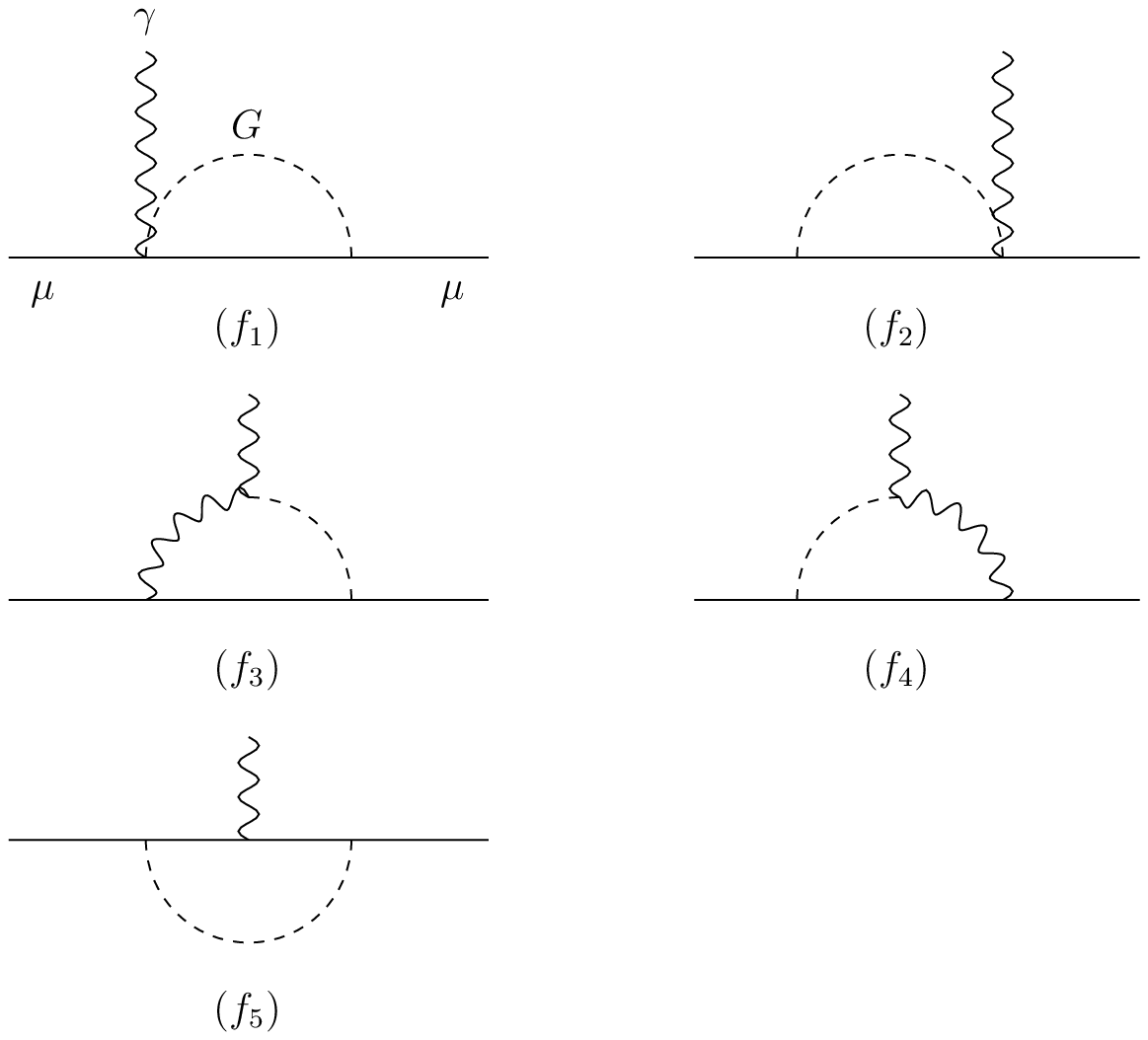,width=10cm,height=8cm}\hss}
\smallskip
\caption{ The Feynman diagrams for the KK graviton
contributions to $\Dt a_\mu$ in the RS model.
}
\label{fig1}
\end{figure}
\end{center}


\begin{center}
\begin{figure}[htb]
\vskip -0.9cm
\hbox to\textwidth{\hss\epsfig{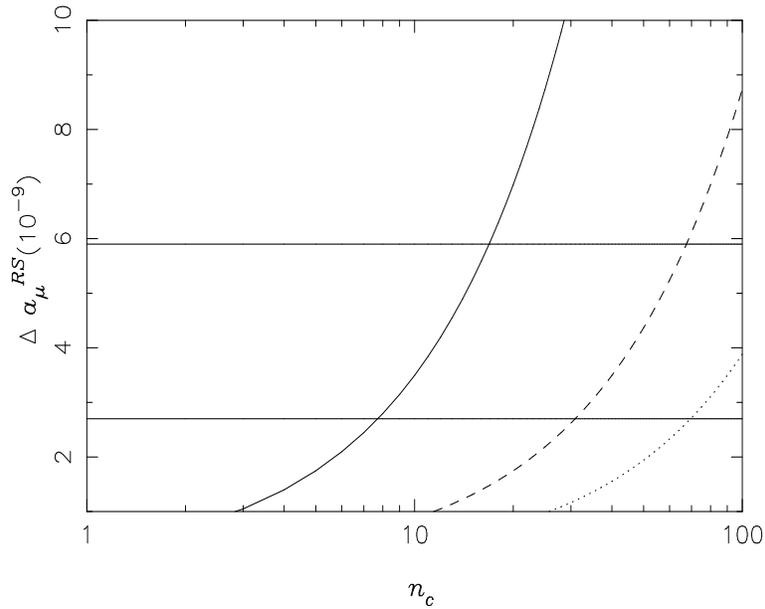}\hss}
\smallskip
\smallskip
\smallskip\smallskip
\caption{ The $\Dt a_\mu$ as a function of $n_c$ for
$\Lm_\pi=1$ (solid curve), 2 (dashed), and 3 (dotted) TeV. The
horizontal lines denote the recent E821 data of $\Dt a_\mu$ within
$1\sigma$ level. } \label{fig2}
\end{figure}
\end{center}

\begin{center}
\begin{figure}[htb]
\hbox
to\textwidth{\hss\epsfig{file=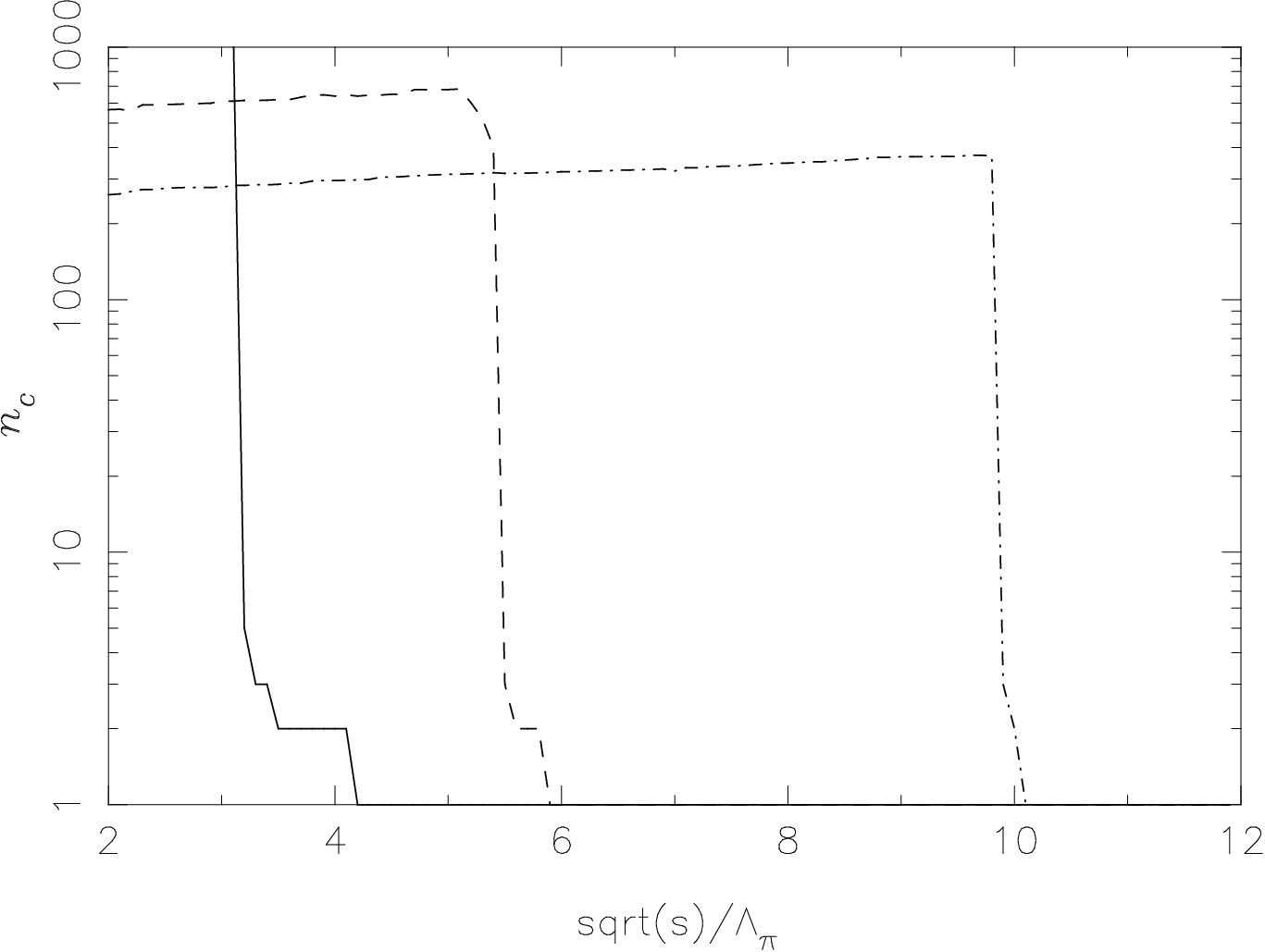,width=8cm,height=10cm,angle=-90}\hss}
 \vskip -.2cm
\smallskip
\smallskip
\smallskip
\caption{ The unitary bounds on $(\sqrt{s}/\Lm_\pi, n_c)$ plane from
the $a^2_{00}$ of the $\gm\gm\to\gm\gm$ process. The $k/M_{\rm
Pl}=0.1$ (solid curve), 0.3 (dashed) and 0.7 (dot-dashed) cases are
considered. } \label{fig3}
\end{figure}
\end{center}

\begin{center}
\begin{figure}[htb]
\hbox
to\textwidth{\hss\epsfig{file=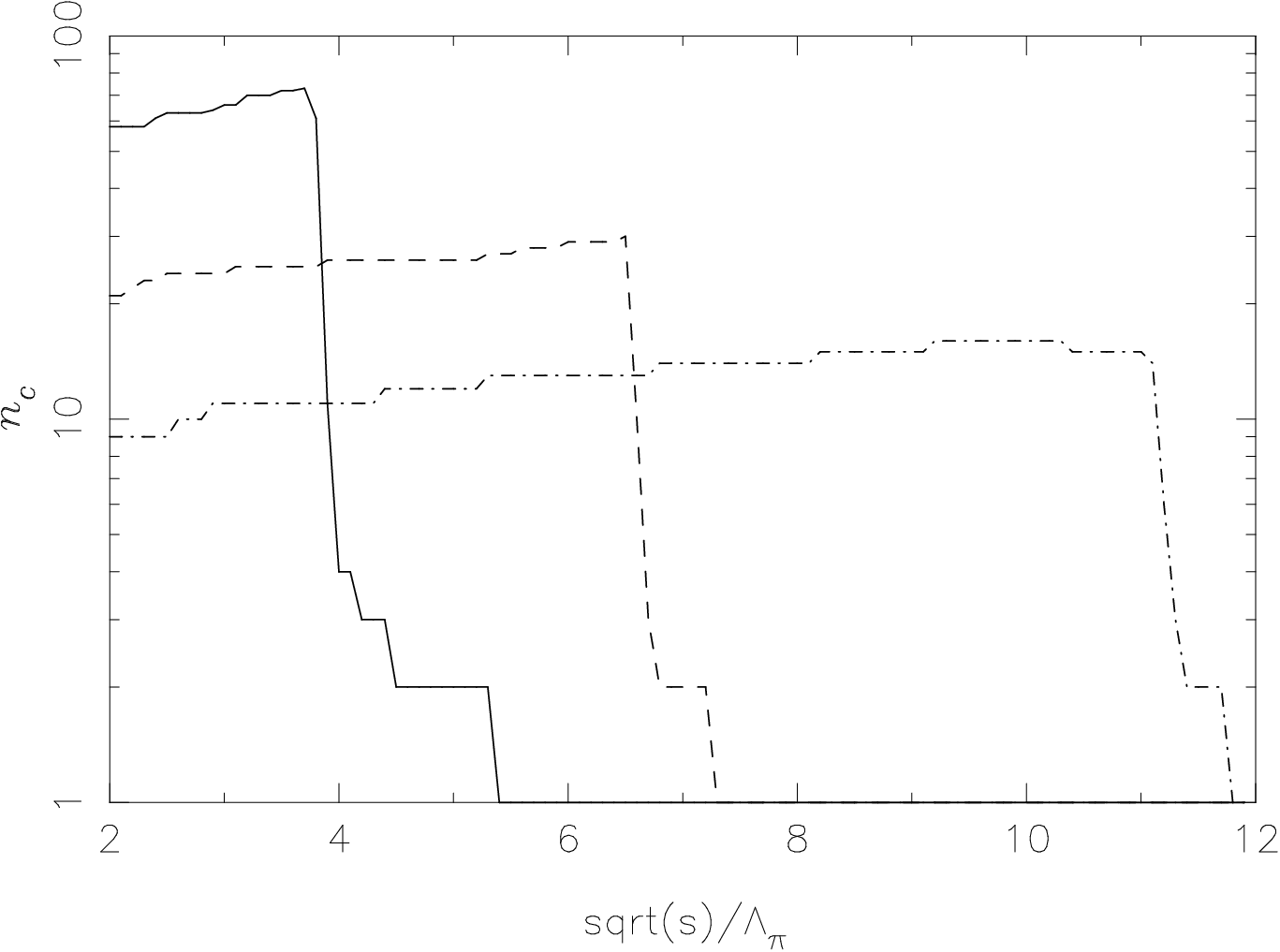,width=8cm,height=10cm,angle=-90}\hss}
\vskip -.2cm
\smallskip
\smallskip
\smallskip
\caption{ The unitary bounds on $(\sqrt{s}/\Lm_\pi, n_c)$ plane from
the $2a^2_{22}$ of the $\gm\gm\to\gm\gm$ process. The $k/M_{\rm
Pl}=0.1$ (solid curve), 0.3 (dashed)  and 0.7 (dot-dashed) cases are
considered. } \label{fig4}
\end{figure}
\end{center}

\end{document}